\documentclass[final,3p,times]{elsarticle}

\usepackage{amssymb}
\usepackage{lipsum}

\usepackage{physics,amsmath,tensor}
\usepackage{mathtools}
\usepackage{pstool}
\usepackage{epstopdf}
\usepackage{psfrag}
\usepackage{graphicx}
\graphicspath{{FIGURES/}}
\usepackage{dcolumn}
\usepackage{bm}
\usepackage[mathlines]{lineno}
\usepackage{xcolor}
\usepackage[normalem]{ulem}
\usepackage{nicefrac,xfrac}

\newcommand{\pati}[1]{}

\newcommand{\ratioAV}{\scriptscriptstyle \!\!{A}\!/\!{V}}

\journal{Annals of Physics}

\begin{document}

\begin{frontmatter}



\title{Access to Klein Tunneling via Space-Time Modulation}



\author{Furkan Ok\corref{cor1}}
\ead{furkan.ok@kuleuven.be}

\author{Amir Bahrami}
\author{Christophe Caloz}

\cortext[cor1]{Corresponding author}

\affiliation{%
  organization={Department of Electrical Engineering, KU Leuven},
  city={Leuven},
  postcode={3001},
  country={Belgium}
}

\begin{abstract}
We show that space-time modulation of electromagnetic potentials enables Klein tunneling far below the static threshold. This effect arises from the oblique transitions inherent in the associated kinematics, which connect opposite-energy continua without requiring their overlap, yielding a velocity-tunable Klein gap in which transmission vanishes over a finite velocity window and reemerges beyond it. The associated reduction in energy thresholds—by up to four orders of magnitude—suggests the possibility of experimental realization using flying-focus fronts and relativistic electron beams.
\end{abstract}




\end{frontmatter}




\section{Introduction}
\pati{Klein Paradox as Long-standing Problem}
In relativistic quantum mechanics, the Dirac equation introduces positive- and negative-energy continua that make scattering at steep electrostatic steps profoundly nonclassical~\cite{Dirac_1928}. Building on Dirac’s formulation, Klein showed that a sufficiently high potential step couples an incident electron to the negative-energy continuum across the Klein gap, leading to an increase in transmission with barrier height---the Klein paradox~\cite{Klein_1929}. This counterintuitive phenomenon has since become a canonical benchmark of relativistic transport, linking high-barrier transmission and vacuum pair creation in strong fields. Comprehensive reviews~\cite{Dombey_1999_Klein,Calogeracos_1999_Klein}, textbooks~\cite{Greiner_1985_QED,Bjorken_1964_RQM} and original works~\cite{Holstein_1998_Klein,Nitta_1999_Motion,Petrillo_2003_Relativistic,Krekora_2004_Klein} provide complementary perspectives.

\pati{Excessive Required Critical Fields}
Shortly after Klein’s discovery, Sauter analyzed the more realistic case of a linear-ramp potential and found that the field required to produce the Klein effect was as high as $\mathcal{E}_\mathrm{c} = m^2 c^3 / |q|\hbar \approx 1.32 \times 10^{18}\,\text{V/m}$---a critical level below which the transition probability to the negative-energy continuum becomes negligible~\cite{Sauter_1931}. This field was later identified by Heisenberg, Euler and Weisskopf~\cite{Heisenberg_1936,Weisskopf_1936} as the onset of electron--positron pair creation, and subsequently expressed by Schwinger~\cite{Schwinger_1951} in terms of the vacuum persistence probability in a constant electric field. The critical field thus defines the characteristic scale for transitions into the negative-energy continuum in vacuum, where the electric force $|q|\mathcal{E}_\mathrm{c}$ performs work equal to the pair rest energy $2mc^2$ over a distance of two reduced Compton wavelengths, $2\hbar/mc$.

\pati{Klein Paradox in Dirac Matter \& Simulations}
The Klein paradox has also been extensively explored in condensed matter systems, most notably in graphene, where low-energy charge carriers behave as massless Dirac fermions~\cite{Abergel_2010_Properties,Cooper_2012_Experimental}. In this case, Klein tunneling can occur without a theoretical threshold field, although disorder tolerance in realistic samples imposes practical requirements on the order of $\mathcal{E}=10^7\,\text{V/m}$~\cite{Katsnelson_2006_Chiral, Beenakker_2008_Colloquium,Allain_2011_Klein}. Experimental observations in graphene-based structures have confirmed \emph{analogue} Klein tunneling for Dirac-like quasiparticles~\cite{Stander_2009_Evidence,Young_2009_Quantum,Elahi_2024_Direct}. Beyond isolated monolayer graphene junctions, Klein-related transport has also been studied in multilayer and spatially patterned graphene, including ABC-stacked trilayer-graphene superlattices and monolayer graphene with superperiodic barrier arrangements~\cite{Saley_2024_Tunneling,Shekhar_2025_Relativistic}. These studies show how interlayer coupling and repeated potential arrangements modify transmission and generate additional gaps and resonances, extending the monolayer picture. More recently, quantum simulation platforms have enabled controlled studies of Klein tunneling, including \mbox{trapped-ion} implementations that emulate the dynamics of Dirac particles~\cite{Casanova_2010_Klein,Gerritsma_2011_Quantum}. These condensed-matter and quantum-simulation realizations emulate Dirac dynamics without producing real electron--positron pairs, and access to the corresponding vacuum Klein regime remains experimentally challenging.

\pati{Attempts to Reduce the Pair Creation Limit}
Several approaches have been proposed to enhance or otherwise control vacuum pair production. In the dynamically assisted Schwinger mechanism, a strong, slowly varying electric field is combined with a weak, rapidly oscillating field, which reduces the tunneling exponent and thereby enhances pair creation. Early work reported a reduction of about $40\%$ in the instanton action under suitable parameters, increasing the predicted yield from roughly one electron--positron pair per year to a few per day~\cite{Schutzhold_2008_Dynamically}. Subsequent studies based on the same mechanism achieved substantially stronger improvements, reaching one- to two-order-of-magnitude enhancements in suitable regimes~\cite{Kohlfurst_2021_Dynamically,Ryndyk_2024_Dynamically}. A complementary approach based on smooth spatial modulation of electromagnetic potentials showed that vacuum tunneling can also be controlled while maintaining strong fields in the interaction region, with long-time pair-creation rates modified by about $40\%$ across various scenarios~\cite{Su_2025_Controlling}. More recently, an analogue of dynamical assistance was studied in the Klein setting, where adding a weak time-periodic field to a static step potential lowered the supercriticality condition and reduced the required step height by approximately $30\%$~\cite{Ochiai_2025_Dynamically}.

\pati{Experimental Challenge in Vacuum}
Despite these theoretical advances, direct observation of Klein tunneling in vacuum remains essentially out of reach because of the extremely high threshold field mentioned above. Recent developments in high-power laser technology have pushed achievable intensities to about $10^{27}\,\mathrm{W/m^{2}}$~\cite{Yoon_2021_Realization}, still several orders of magnitude below the $10^{33}\,\mathrm{W/m^{2}}$ corresponding to the Schwinger critical field.

\pati{Here: Space-Time Modulation Solution}
In this work, we study Klein (non-evanescent) tunneling in the presence of \emph{space--time} engineered step potentials, showing that velocity modulation introduces a new degree of freedom for controlling transitions into the negative-energy continuum. From the viewpoint of classical electromagnetic metamaterials, space-time modulation has long been recognized as a powerful tool, beginning with early studies of periodic structures and moving interfaces~\cite{Cassedy_PI_1963,Bolotovskiui_SPU_1972,Lampert_PR_1956,Balazs_JMAA_1961,Semenova_RQE_1972,Yablonovitch_PRL_1974,Ostrovski_SPU_1975}, and has evolved into a comprehensive framework for wave control~\cite{Caloz_2019_Spacetime1,Caloz_2019_Spacetime2,Engheta_2021_Metamat,Huidobro_PRApp_2021,Galiffi_NC_2022,Li_PRApp_2023,Liu_OE_2024,Wang_PRL_2025,Pacheco_Pena_PRB_2025}. While these concepts are well established in classical systems, their application to quantum electronic systems remains comparatively unexplored. Early demonstrations include magnet-free nonreciprocal plasmonic devices based on conductivity modulation in graphene~\cite{Correas_AWPL_2016}, the classification of space-time groups with topological properties~\cite{Xu_PRL_2018}, oblique spacetime crystals exhibiting energy conversion via Floquet-Bloch oscillations~\cite{Gao_PRL_2021,Gao_PRB_2022}, and the topological classification of spacetime crystals using a frequency-domain-enlarged Hamiltonian~\cite{Peng_PRL_2022}.

\pati{Contribution}
Here we show that space-time modulation of the electromagnetic potentials produces oblique energy--momentum transitions that (i)~enable access to Klein tunneling at substantially reduced scalar steps---without overlap between the positive- and negative-energy continua, (ii)~render the Klein gap continuously tunable via the modulation velocity and vector-scalar-potential offsets, and (iii)~yield a velocity-dependent Klein paradox in which transmission vanishes within a finite velocity window and reemerges beyond this window. We analyze a single subluminal modulation in $(1+1)$ dimensions, derive closed-form lab-frame kinematics and scattering amplitudes, and evaluate probabilities from the Dirac current projected on the modulation worldline.

\section{Dirac Equation}
\pati{Dirac Equation}
In natural units $(\hbar=c=1)$, the minimally coupled 1+1D Dirac Hamiltonian reads~\cite{Dirac_1928,Greiner_1985_QED,Bjorken_1964_RQM,Peskin_2018_QFT}
\begin{equation}\label{eq:Dirac_H_1p1D}
    H=\sigma_x\left(-i\,\partial_z - qA\right)+\sigma_z\,m+qV,
\end{equation}
where $\sigma_{x,y,z}$ are Pauli matrices, $m$ is the electron rest mass, $V(t,z)$ is the scalar potential, $A(t,z)$ is the vector potential, and we adopt the Minkowski metric $\eta^{\mu\nu}=\mathrm{diag}(1,-1)$. Inserting the positive-energy monochromatic traveling plane-wave ansatz $\psi(t,z)~\!=~\!u(p)\,\mathrm{e}^{-i(Et-pz)}$ into Eq.~\eqref{eq:Dirac_H_1p1D} leads to the general spinor solution form (see Sec.~1 in~\cite{supp_mat})
\begin{equation}\label{eq:plane_wave_sol}
    \psi=
        \begin{pmatrix}
        1\\
        \frac{E-qV-m}{p-qA}
        \end{pmatrix}
    \mathrm{e}^{-i(E t - p z)},
\end{equation}
where $q = -e$ ($e > 0$) is the charge of the electron.

\section{Conventional Klein paradox}
\pati{Static Klein Step Scattering Coefficients}
For a static (purely spatial) scalar potential interface $(A = 0)$ located at $z = z_{0}$, energy is conserved while the momentum may change. Applying the plane-wave spinors from Eq.~\eqref{eq:plane_wave_sol} on each side, continuity of the spinor wave function at $z_{0}$ determines the reflection and transmission amplitudes. The corresponding probabilities follow from the normal component of the Dirac current, $j^{z} = \vb{j}\cdot\vb{\hat{n}}$, with $\vb{\hat{n}} = \vb{\hat{z}}$ the unit normal, ensuring flux continuity among the incident, reflected and transmitted channels~\cite{Greiner_1985_QED,Bjorken_1964_RQM}.

\begin{figure}[ht!]
    \centering
    \includegraphics[width=0.65\textwidth]{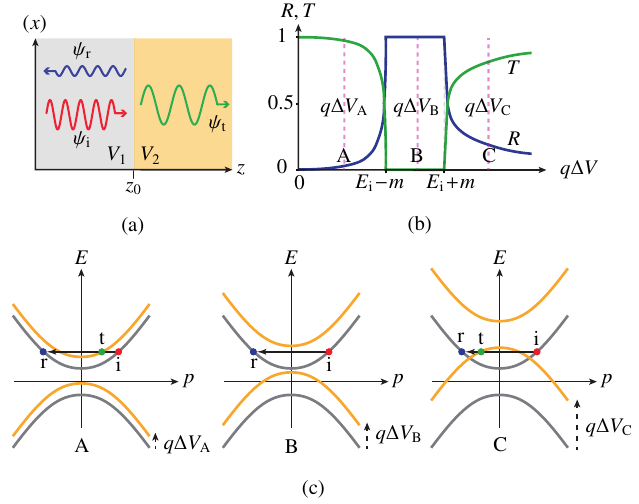}
    \vspace{-3mm}
    \caption{Conventional Klein paradox at a spatial scalar potential step. (a)~Single interface at $z_0$ between $V_1$ and $V_2$ (we set $V_1=0$) with incident wave $\psi_\mathrm{i}$, reflected wave $\psi_\mathrm{r}$ and transmitted wave $\psi_\mathrm{t}$. (b)~Current-ratio probabilities $R=-j_\mathrm{r}/j_\mathrm{i}$ and $T=j_\mathrm{t}/j_\mathrm{i}$ versus step height $q\Delta V$ at fixed $E_\mathrm{i}$, with subcritical region $q\Delta V_\mathrm{A}<E_\mathrm{i}-m$, Klein gap $E_\mathrm{i}-m<q\Delta V_\mathrm{B}<E_\mathrm{i}+m$ (evanescent $\psi_\mathrm{t}$) and supercritical or Klein-region $q\Delta V_\mathrm{C}>E_\mathrm{i}+m$ (coupling to the negative-energy branch). (c)~Corresponding dispersion diagrams, with $\mathrm{i},\mathrm{r},\mathrm{t}$ denoting incidence, reflection and transmission, respectively, and where the last panel corresponds to conventional Klein tunneling.}\label{fig:ST_Conven}
\end{figure}

\pati{Continua Overlap Challenge}
Figure~\ref{fig:ST_Conven} illustrates the standard case of a purely spatial scalar potential step and three complementary views of the problem: (a)~a direct-space schematic, (b)~reflection and transmission probabilities versus step height, and (c)~dispersion $(p,E)$ diagrams. As the step height increases [Figs.~\ref{fig:ST_Conven}(b) and (c)], the response evolves from ordinary partial reflection (subcritical region---A) to a region with no propagating transmitted mode (the Klein gap---B) and finally to transition into the negative-energy continuum (supercritical region---C), where transmission increases with step height---the hallmark of the Klein paradox~\cite{Klein_1929,Dombey_1999_Klein,Calogeracos_1999_Klein,Greiner_1985_QED,Bjorken_1964_RQM,Holstein_1998_Klein,Nitta_1999_Motion,Petrillo_2003_Relativistic}. Pair creation arises in this energy range, the so-called Klein-region~\cite{Longhi_2010_klein,Lv_2013_quantum,Su_2025_Controlling,Ochiai_2025_Dynamically}, where the positive- and negative-energy continua overlap, and whose limit, $E_\mathrm{i}+m$, may approach the minimum $2m$, corresponding to $\mathcal{E}_\mathrm{c}$.

\section{Time- and Space-Time Modulations}
\pati{Pure-Time Modulation Recall: No Gap Crossing}
For a purely temporal modulation—the infinite-velocity limit of superluminal space-time modulation~\cite{Caloz_2019_Spacetime2}—the potential is uniform in space and varies only in time. Whereas spatial modulation implies a momentum transformation with conserved energy due to temporal translational symmetry~\cite{Noether_1918}, temporal modulation breaks that symmetry and therefore induces an energy transformation while conserving momentum~\cite{Mendoncca_2021_temporal,Kim_2023_propagation}. In this regime, transmission into the later medium occurs regardless of the potentials, since only that medium exists after the transition. In terms of the dispersion diagram, the transition line, being vertical, necessarily intersects the positive-energy branch of the second medium~\cite{Ok_SR_2024}. Therefore, the notion of a Klein gap becomes immaterial. The Klein gap arises only when the modulation is \emph{subluminal}, a space-time regime in which a comoving frame exists where energy is conserved and an evanescent transmitted channel—the Klein gap—can emerge.

\begin{figure}[ht!]
    \centering
    \includegraphics[width=0.65\textwidth]{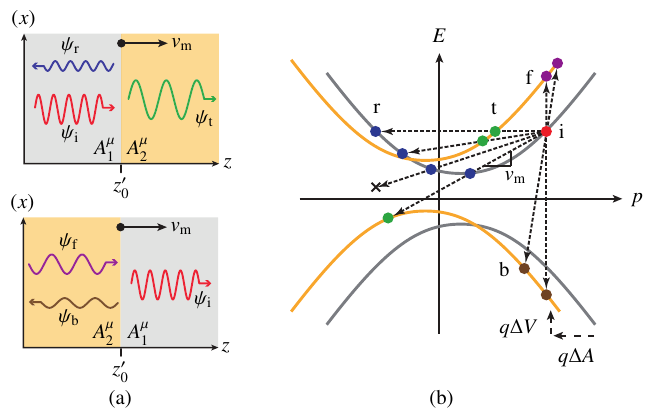}
    \vspace{-3mm}
    \caption{Scattering at a space-time modulation interface. (a)~Direct space schematic, with an interface between the four-potentials $A_1^\mu$ and $A_2^\mu$ moving at velocity $v_\mathrm{m}$ (top: subluminal; bottom: superluminal) and channels: i (incident), r (reflected), t (transmitted), b (later-backward), f (later-forward). (b)~Dispersion diagram with transitions: horizontal = pure-space, vertical = pure-time, oblique = space-time (sub/superluminal). Vertical/horizontal offsets correspond to scalar/vector steps $q\Delta V$ and $q\Delta A$.}
\label{fig:ST_step_and_disp}
\end{figure}

\pati{Introduction of Space-Time Modulation}
Figure~\ref{fig:ST_step_and_disp} illustrates a general space-time engineered modulation step, shown both in real space [Fig.~\ref{fig:ST_step_and_disp}(a)] and in the dispersion diagram [Fig.~\ref{fig:ST_step_and_disp}(b)]. A modulation interface moving at velocity $v_\mathrm{m}$ preserves a well-defined combination of energy and momentum determined by Lorentz transformations, producing transitions with slope $v_\mathrm{m}$ in the dispersion diagram. The transitions are horizontal in the pure-space limit ($v_\mathrm{m}\!=\!0$), vertical in the pure-time limit ($v_\mathrm{m}\!\to\!\infty$) and oblique for finite velocities, with slopes less than unity ($v_\mathrm{m}\!<\!1$) corresponding to the subluminal regime and slopes greater than unity ($v_\mathrm{m}\!>\!1$) to the superluminal regime.

\pati{Subluminal Constraints}
In the subluminal regime, scattering occurs when the incident wave overtakes the interface [Fig.~\ref{fig:ST_step_and_disp}(a), top], whereas in the superluminal regime, the wave cannot catch up with the interface, so nontrivial scattering requires placing the incident wave in the second medium [Fig.~\ref{fig:ST_step_and_disp}(a), bottom]. However, this configuration does not lead to a Klein gap because the associated transitions have slopes greater than $45^{\circ}$, forcing them to intersect the positive-energy continuum and thereby enabling transmission, consistently with the absence of an energy gap in the comoving frame. Therefore, the Klein gap can occur only in the subluminal regime, to which we restrict our analysis.

\pati{Codirectional Constraints}
Moreover, the \emph{codirectional regime} ($v_\mathrm{m}>0$), in which the modulation travels in the same direction as the incident electron wave, is far more favorable than the contradirectional regime ($v_\mathrm{m}<0$), in which the modulation and the wave propagate in opposite directions. Indeed, in the former case, the transitions have positive slopes and are directed toward the lower left of the diagram, as illustrated in Fig.~\ref{fig:ST_step_and_disp}(b), implying a reduction in the required potential compared with the conventional case (see Fig.~\ref{fig:ST_Conven}). In contrast, in the contradirectional regime, the transitions have negative slopes and point toward the upper left of the dispersion diagram, leading to Klein energies even higher than in the conventional case.

\pati{Moving modulation versus moving matter}
It is crucial to keep in mind that the present interface is a \emph{moving modulation}, not moving matter. In the electromagnetic case, moving-matter media become bianisotropic in the laboratory frame~\cite{Kong_EWT_2008} because the motion of the constituent matter induces magnetoelectric coupling~\cite{rontgen1888ueber}, whereas moving-modulation media involve no net transport of atoms or molecules: only the externally driven perturbation moves, yet the resulting effective medium also acquires a cross-coupled constitutive structure~\cite{Caloz_APM_12_2022}. Similarly, the present Dirac problem, which involves a traveling modulation rather than transported matter, constitutes a \emph{mixed vector-scalar potential} problem analogous to electromagnetic bianisotropy and is therefore fundamentally distinct from static matter scattering.

\pati{Energy-Momentum Transitions}
In each uniform region, the spectrum satisfies the Dirac dispersion $(E - qV)^2 = (p - qA)^2 + m^2$. To describe a modulation moving at velocity $v_\mathrm{m}$, we pass to the comoving frame (primes). By Noether’s theorem, time-translation symmetry in that frame implies energy conservation, $E^{\prime}_\mathrm{i}=E^{\prime}_\mathrm{r}=E^{\prime}_\mathrm{t}$, and the channel energies are related to each other by the Lorentz transformation $E_a^{\prime}=\gamma_\mathrm{m}(E_a - v_\mathrm{m}p_a)$, ($a=\mathrm{i},\mathrm{r},\mathrm{t}$), where $\gamma_\mathrm{m} = (1 - v_\mathrm{m}^{2})^{-1/2}$. Solving these relations together with the dispersion gives the laboratory-frame energies and momenta of the reflected and transmitted waves in terms of the input values (see Sec.~2 in~\cite{supp_mat}),
\begin{subequations}\label{eq:lab_var}
    \begin{equation}
    E_\mathrm{r} = E_\mathrm{i} + 2\gamma_\mathrm{m}^2 v_\mathrm{m}  W_\mathrm{r},
    \end{equation}
    \begin{equation}
    p_\mathrm{r}=p_\mathrm{i} + 2\gamma_\mathrm{m}^{2} W_\mathrm{r},
    \end{equation}
    \text{where} $W_\mathrm{r} = v_\mathrm{m} (E_\mathrm{i}-qV_1) - (p_\mathrm{i}-qA_1)$, \text{and}
    \begin{equation}
    E_\mathrm{t}^{\pm} = \gamma_\mathrm{m}^2 W_\mathrm{t} \pm \gamma_\mathrm{m}^2 v_\mathrm{m} \sqrt{W_\mathrm{t}^2 - \left(m/\gamma_\mathrm{m}\right)^2} + qV_2,
    \end{equation}
    \begin{equation}\label{eq:pt}
    p_\mathrm{t}^{\pm} = \gamma_\mathrm{m}^{2} v_\mathrm{m} W_\mathrm{t} \pm \gamma_\mathrm{m}^{2} \sqrt{W_\mathrm{t}^2 - \left(m/\gamma_\mathrm{m}\right)^2}  + qA_2,
\end{equation}
\end{subequations}
where $W_\mathrm{t} = (E_\mathrm{i}-qV_2) - v_\mathrm{m} \left(p_\mathrm{i} - qA_2\right)$.
The signs of the transmission energies and momenta are chosen to ensure physically admissible wave solutions~\cite{Klein_1929,Greiner_2000_RQM} (see Sec.~3 in~\cite{supp_mat}).

\pati{Scattering Amplitudes of Oblique Transitions}
Matching is conveniently performed in the comoving frame, where the interface is stationary and the transformed spinor wave function is continuous at the boundary, $\left.\psi^\prime_1\right|_{z^\prime=z^\prime_{0}}  =  \left.\psi^\prime_2\right|_{z^\prime=z^\prime_{0}}$, after applying a suitable spinor boost $S_z(\Lambda)\psi(t,z)=\psi^{\prime}(t^{\prime},z^{\prime})$ (see Sec.~4 in~\cite{supp_mat}). Because, for a uniformly moving interface, $S_z(\Lambda)$ is a constant invertible linear transformation, this continuity condition is equivalent to continuity of the spinor in the laboratory frame when evaluated on the moving interface worldline $z-v_{\mathrm m}t=z_0$. Using plane-wave spinors on each side together with the laboratory-frame channel variables obtained in Eqs.~\eqref{eq:lab_var}, this continuity condition reduces to a $2\times2$ linear system for the unknown amplitudes. Solving it yields the reflection and transmission amplitudes (see Sec.~5.1 in~\cite{supp_mat})
\begin{equation}
    r = \frac{\Gamma_\mathrm{i}-\Gamma_\mathrm{t}}{\Gamma_\mathrm{t}-\Gamma_\mathrm{r}}
    \quad \text { and } \quad
    t = \frac{\Gamma_\mathrm{i}-\Gamma_\mathrm{r}}{\Gamma_\mathrm{t}-\Gamma_\mathrm{r}},
\end{equation}
where $\Gamma_\mathrm{i}   =   \frac{E_\mathrm{i}-qV_1-m}{p_\mathrm{i}-qA_1}$, $\Gamma_\mathrm{r}   =   \frac{E_\mathrm{r}-qV_1-m}{p_\mathrm{r}-qA_1}$ and $\Gamma_\mathrm{t}   =   \frac{E_\mathrm{t}-qV_2-m}{p_\mathrm{t}-qA_2}$.

\pati{Probability Calculations}
In the purely spatial case, reflection and transmission probabilities are obtained from the flux of the Dirac current through the modulation interface plane. The same principle applies to a space-time modulation, whose level set is described as $f(t,z) := z - v_\mathrm{m}t = z_{0}$. We take the (unnormalized) covector normal to the modulation as $n_{\mu}\propto \partial_{\mu}f  =  (-v_\mathrm{m},1)$. The flux carried by each channel is given by the projection of its current $j^{\,\mu} = (\rho,j^{\,z})$ onto this normal, $j_{a} = j^{\,\mu}_an_{\mu} = j^{\,z}_a-v_\mathrm{m}\rho_a$ ($a=\mathrm{i},\mathrm{r},\mathrm{t}$), where $\rho \equiv j^{0}$ is the charge (probability) density and $j^{z}$ is the longitudinal current. The reflection and transmission probabilities are then obtained as~\cite{Auletta_2009_quantum}
\begin{equation} \label{eq:probability}
    R=-j_{\mathrm r}/j_{\mathrm i}\quad \text{and} \quad T=j_{\mathrm t}/j_{\mathrm i},
\end{equation}
ensuring $R+T=1$ (see Sec.~5.2 in~\cite{supp_mat}). In the spatial limit $v_\mathrm{m} \to 0$, this reduces to the usual surface-normal current through $z = \text{const}$, i.e.\ $j^{z}$.

\section{Reduced threshold}
\pati{Potential Threshold Equation}
Within the Klein gap, the square root term in Eq.~\eqref{eq:pt} is negative, making $p_{\mathrm t}$ complex and the transmitted mode is evanescent. The gap edges are obtained by setting that root to zero, corresponding to the transition between real and evanescent $p_{\mathrm t}$. In the dispersion diagram, these edge conditions correspond to the tangency points of the transition line from the incident state [red point in Fig.~\ref{fig:ST_step_and_disp}(b)] to the top and bottom dispersion branches for the second medium [brown curves in Fig.~\ref{fig:ST_step_and_disp}(b)], associated with $q\Delta V^{+}$ and $q\Delta V^{-}$. These energy offsets are given by
\begin{equation} \label{eq:threshold}
    q\Delta V^{\pm} = \frac{(E_\mathrm{i}-qV_1) - v_\mathrm{m} (p_\mathrm{i}-qA_1) \mp m/\gamma_\mathrm{m}}{1 - v_\mathrm{m} r_{\ratioAV}},
\end{equation}
where $r_{\ratioAV}\equiv\Delta A^{\pm}/\Delta V^{\pm}$, and $q\Delta V^{-}$ coincides with the threshold potential $q\Delta V^{\mathrm{th}}$ of Klein tunneling (see Sec.~6.1 in~\cite{supp_mat}). For the gap to occur, one must also avoid the case where the transition line intersects the second-medium upper dispersion branch, which would open a propagating channel into the positive-energy continuum. The corresponding condition requires the vector-potential offset to be smaller than the scalar step, $q\Delta A<q\Delta V$. In the forthcoming results, we shall take for definiteness $q\Delta V>0$ (upward scalar step), reducing the gap condition to $r_{\ratioAV}<1$.

\begin{figure}[ht!]
    \centering
    \includegraphics[width=0.85\textwidth]{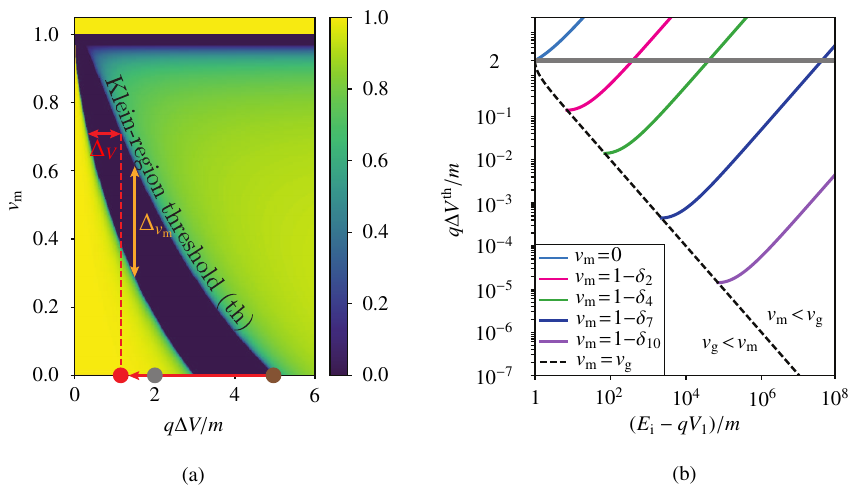}
    \vspace{-3mm}
    \caption{Klein-region response to a subluminal space-time modulation step with $r_{\ratioAV}\!=\!-1$.
    (a)~Transmission probability $T$ for an electron with $E_{\mathrm i}-qV_1\!=\!4m$, versus $q\Delta V/m$ and $v_{\mathrm m}$. Brown point: static threshold $(E_{\mathrm i}\!-\!qV_1\!+\!m)/m$; gray point: minimum static threshold $q\Delta V^{\mathrm{th}}_{\mathrm{static,min}}/m\!=\!2$; red point: dynamic threshold.
    (b)~Normalized Klein-region threshold $q\Delta V^{\mathrm{th}}/m$ for different values of $v_{\mathrm m}$ near the speed of light, corresponding to higher electron energies, where the Klein gap is too narrow to be visible on the scale of (a). Solid curves: $v_{\mathrm m}\!=\!1\!-\!\delta_n$, $\delta_n\!=\!10^{-n}$; dashed curve: velocity-matching limit $v_{\mathrm m}\!=\!v_{\mathrm g}$; horizontal gray line: minimum static threshold $q\Delta V^{\mathrm{th}}_{\mathrm{static,min}}/m\!=\!2$.}\label{fig:ST_probability}
\end{figure}

\pati{Potential Threshold Reduction at All $v_\text{m}$}
Figure~\ref{fig:ST_probability} shows how the space-time modulation step of the electromagnetic potentials controls the transmission and the threshold potential. Figure~\ref{fig:ST_probability}(a), computed from Eq.~\eqref{eq:probability}, plots the transmission for the electron with $E_{\mathrm i}-qV_1=4m$, as a function of $q\Delta V/m$ and $v_\mathrm{m}$. The dark curved wedge is the \emph{generalized Klein gap}, where the transmitted momentum is evanescent. As $v_\mathrm{m}$ increases, the energy gap closes and the Klein-region threshold (right edge of the wedge)---the \emph{dynamic} energy threshold---$q\Delta V^{\mathrm{th}}$ shifts to values both drastically below the static threshold for $E_{\mathrm i}-qV_1=4m$ (case of the figure) and even well below the minimum reference $q\Delta V_{\mathrm{static,min}}^{\mathrm{th}}=2m$, corresponding to the Schwinger critical limit in terms of the electric field. The horizontal dark-blue band corresponds to $v_\mathrm{g}<v_\mathrm{m}<1$, with $v_\mathrm{g}$ being the group velocity of the incident electron, where the electron cannot catch up with the modulation\footnote{The thin yellow superluminal ($v_\mathrm{m}>1$) band at the top marks the speed of light boundary ($v_\mathrm{m}=1$) of the no-catch-up regime, beyond which the system enters the region shown at the bottom of Fig.~\ref{fig:ST_step_and_disp}, where no Klein regime exists.}.

\pati{Potential Threshold Reduction at $v_\text{m}\approx{c}$}
Figure~\ref{fig:ST_probability}(b), obtained from Eq.~\eqref{eq:threshold}, plots $q\Delta V^{\mathrm{th}}/m$ versus $(E_{\mathrm i}-qV_1)/m$ for fixed $v_\mathrm{m}$. The colored solid curves correspond to $v_\mathrm{m}=1-\delta_n$ with $\delta_n=10^{-n}$ ($n=2,4,7,10$), while the dashed line marks the limit $v_\mathrm{m}=v_\mathrm{g}$. No curve appears to the left of this dashed limit because in that domain $v_\mathrm{g}<v_\mathrm{m}$ and the electron cannot catch up with the modulation front. In the physically relevant region, where $v_\mathrm{m}<v_\mathrm{g}$ (to the right of the dashed limit), the threshold drops by orders of magnitude as $v_\mathrm{m}\!\to\!1^{-}$, reaching its minimum near $v_\mathrm{m}=v_\mathrm{g}$; along this limit, the threshold collapses, i.e., $q\Delta V^{\mathrm{th}}\!\to\!0$. Representative minima are $1.4\times10^{-1}m$ for $v_\mathrm{m}=1-10^{-2}$, $1.4\times10^{-2}m$ for $v_\mathrm{m}=1-10^{-4}$, $4.5\times10^{-4}m$ for $v_\mathrm{m}=1-10^{-7}$, and $1.4\times10^{-5}m$ for $v_\mathrm{m}=1-10^{-10}$. We display only cases with $\delta_n\!\ge\!10^{-10}$; smaller $\delta_n$ values continue the trend but add little visual information while requiring prohibitively precise velocity matching ($v_\mathrm{m}\!\to\!v_\mathrm{g}$). The curve for $v_\mathrm{m}=0$ recovers the static case. For all curves shown, we set $r_{\ratioAV}=-1$ (see Fig.~\ref{fig:ST_step_and_disp}), which corresponds to scalar and vector potential offsets having equal magnitude and opposite signs. In practice, the magnitude of this ratio can naturally range from zero (a purely scalar potential $V$) to infinity (a purely vector potential $A$).

\pati{Field Threshold Reduction}
Along the velocity-matching limit ($v_{\mathrm m}\approx v_{\mathrm g}$), the required scalar step is reduced to $q\Delta V^{\mathrm{th}}=2m\,e^{-\omega_{\mathrm g}}$, with electron rapidity $\omega_{\mathrm g}=\operatorname{arctanh}(v_{\mathrm g})$ (see Sec.~6.2 in~\cite{supp_mat}). For traveling longitudinal potentials $V(\xi)$ and $A(\xi)$, with $\xi=z-v_{\mathrm m}t$, the laboratory-frame electric field is $\mathcal E=-\partial_z V-\partial_t A=-\,d\left(V-v_{\mathrm m}A\right)/d\xi$, so that a variation of the combination $V-v_{\mathrm m}A$ over a lab-measured thickness $L$ gives $|\mathcal E^{\mathrm{th}}|\simeq\left|\Delta V^{\mathrm{th}}-v_{\mathrm m}\Delta A^{\mathrm{th}}\right|/L$. At fixed incident state and fixed $v_{\mathrm m}$, the threshold value of this combination---and hence the required physical field---is independent of $r_{\ratioAV}$; the ratio $r_{\ratioAV}$ only redistributes the requirement between the separate scalar and vector potential offsets. At velocity matching, $|q|\left|\Delta V^{\mathrm{th}}-v_{\mathrm g}\Delta A^{\mathrm{th}}\right|=2m(1+v_{\mathrm{g}})e^{-\omega_{\mathrm{g}}}$, hence
\begin{equation}
    \frac{|\mathcal{E}^{\mathrm{th}}|}{\mathcal{E}_{\mathrm{c}}}
    \simeq
    \frac{|q|\,|\Delta V^{\mathrm{th}}-v_{\mathrm{g}}\Delta A^{\mathrm{th}}|}{m^2L}
    =
    2(1+v_{\mathrm{g}})e^{-\omega_{\mathrm{g}}}\frac{\lambda_{\mathrm{C}}}{L},
\end{equation}
with $\mathcal{E}_\mathrm{c}=m^2/|q|$, the reduced Compton
wavelength $\lambda_{\mathrm C}=1/m$, and the Lorentz factor associated with the incident-electron group velocity $\gamma_{\mathrm g}\equiv(1-v_{\mathrm g}^{2})^{-1/2}$, where
$(1+v_{\mathrm g})e^{-\omega_{\mathrm g}}=1/\gamma_{\mathrm g}$. For a non-abrupt yet nonadiabatic modulation step, the interface must
vary on a scale comparable to or smaller than the electron's reduced Compton wavelength~\cite{Sauter_1931,Gies_2005_pair,
Christillin_2007_role,Ruffini_2010_electron}. As a representative case, for $\gamma_{\mathrm g}=10^{4}$ one has
$e^{-\omega_{\mathrm g}}\approx5\times10^{-5}$; with
$L=\lambda_{\mathrm C}$ this yields $|\mathcal{E}^{\mathrm{th}}|/\mathcal{E}_\mathrm{c} \approx 2.0\times10^{-4}$. The scaling is linear in $\lambda_{\mathrm C}/L$: $L=0.5\,\lambda_{\mathrm C}\Rightarrow$ $|\mathcal E^{\mathrm{th}}|/\mathcal{E}_\mathrm{c} \approx 4.0\times10^{-4}$, whereas $L=2\,\lambda_{\mathrm C}\Rightarrow$ $|\mathcal E^{\mathrm{th}}|/\mathcal{E}_\mathrm{c} \approx 1.0\times10^{-4}$. These fields remain far below the static critical value ($\mathcal{E}_\mathrm{c} \simeq 1.3\times10^{18}\,\mathrm{V/m}$) and lie within the reach of present high-intensity lasers ($\sim 10^{26}$--$10^{27}\,\mathrm{W/m^2}$) and laser-plasma structures~\cite{Yoon_2021_Realization}. Flying-focus pulses can program near-luminal ionization fronts with $v_{\mathrm m}\to1^{-}$, and even superluminal fronts in related regimes~\cite{Turnbull_PRL_2018,Sainte_Optica_2017,Ambat_OE_2023}, while modern wakefield and beam-driven accelerators provide electrons with $v_{\mathrm g} \approx 1$~\cite{Miao_PRX_2022,Esarey_RMP_2009,Blumenfeld_Nature_2007,Litos_Nature_2014}.

\pati{Velocity-Mismatch Sensitivity}
The threshold dependence on the modulation velocity can be formulated analytically. Assuming $r_{\ratioAV}=-1$ and defining the
electron and modulation rapidities as
$\omega_{\mathrm g}=\operatorname{arctanh}(v_{\mathrm g})$ and
$\omega_{\mathrm m}=\operatorname{arctanh}(v_{\mathrm m})$, respectively, the
Klein-tunneling threshold, given in Eq.~\eqref{eq:threshold}, can be exactly written as
(see Sec.~6.3.1 in~\cite{supp_mat})
\begin{equation}
\label{eq:mismatch}
    q\Delta V^{\mathrm{th}}(v_{\mathrm m})
    =
    2m\,e^{-\omega_{\mathrm m}}
    \cosh^{2}\!\left(
        \frac{\omega_{\mathrm g}-\omega_{\mathrm m}}{2}
    \right).
\end{equation}
In the velocity-matching limit
$v_{\mathrm m}\to v_{\mathrm g}^{-}$, this reduces to
$q\Delta V^{\mathrm{th}}(v_{\mathrm g})=2m\,e^{-\omega_{\mathrm g}}$,
whereas at $v_{\mathrm m}=0$ it recovers the static threshold
$E_{\mathrm i}-qV_1+m$. The fact that the electron must overtake the
modulation front implies that
$v_{\mathrm m}<v_{\mathrm g}$. Defining then
$\delta v\equiv v_{\mathrm g}-v_{\mathrm m}>0$, the fractional increase
of the threshold near velocity matching is
\begin{equation}
\label{eq:mismatch_expansion}
    \frac{
        \Delta V^{\mathrm{th}}(v_{\mathrm m})
        -
        \Delta V^{\mathrm{th}}(v_{\mathrm g})
    }{
        \Delta V^{\mathrm{th}}(v_{\mathrm g})
    }
    \simeq
    \gamma_{\mathrm g}^{2}\delta v,
\end{equation}
provided that $\gamma_{\mathrm g}^{2}\delta v\ll1$ (see Sec.~6.3.2 in~\cite{supp_mat}). Hence, denoting by $\epsilon$ the tolerated fractional increase of the threshold above its velocity-matched value $\Delta V^{\mathrm{th}}(v_{\mathrm g})$, the required matching precision is approximately $\delta v\lesssim\epsilon/\gamma_{\mathrm g}^{2}$. For $\epsilon=0.1$, corresponding to a representative increase of $10\%$, this gives approximate tolerances of $\delta v\lesssim10^{-5}$, $10^{-7}$ and $10^{-9}$ for $\gamma_{\mathrm g}=10^{2}$, $10^{3}$, and $10^{4}$, respectively. Thus, achieving a four-order-of-magnitude threshold reduction requires a very precise velocity matching.

\pati{Sharp versus Smooth Modulation Front} 
The threshold $q\Delta V^{\mathrm{th}}=2m\,e^{-\omega_{\mathrm g}}$ is a kinematic channel-opening condition determined by the constant potential values far from the front and by the velocity-matching condition. Therefore, for a uniformly traveling smooth front connecting $(V_1,A_1)$ and $(V_2,A_2)$ on its two sides, the $e^{-\omega_{\mathrm g}}$ threshold scaling is unchanged. The front width and profile instead affect the transition probability. As the front width becomes large compared with the reduced Compton wavelength $\lambda_{\mathrm C}$ introduced above, the evolution becomes increasingly adiabatic, and conversion into the negative-energy branch is expected to be suppressed~\cite{Sauter_1931,Gies_2005_pair, Christillin_2007_role,Ruffini_2010_electron}. Consequently, even when the lower-branch energy channel is kinematically open, the transmission may remain small and increase only gradually as the modulation strength is increased.

\pati{Experimental Realization} 
One possible realization of the proposed dynamic Klein scheme would be to use a relativistic electron bunch~\cite{Miao_PRX_2022} co-propagating with a programmable near-luminal ionization front generated by flying-focus techniques~\cite{Turnbull_PRL_2018,Sainte_Optica_2017,Ambat_OE_2023}. The front velocity would be chosen slightly below the electron velocity so that the electrons overtake the modulation. The main experimental requirements are accurate velocity matching and synchronization within the tolerances quoted above, generation of a sufficiently strong and stable traveling field profile, and a sufficiently nonadiabatic front to avoid strong suppression of interbranch transitions. The transmitted population could be characterized by measuring its downstream energy or momentum distribution.

\section{Gap engineering}
\pati{Engineering of the Klein Gap}
The Klein energy gap---horizontal span in Fig.~\ref{fig:ST_probability}(a)---follows from Eq.~\eqref{eq:threshold} as
\begin{equation}
\Delta_V \equiv q\Delta V^{-}-q\Delta V^{+}
= \frac{2m/\gamma_{\mathrm m}}{1-v_\mathrm{m}r_{\ratioAV}}\,,
\end{equation}
valid for $r_{\ratioAV} < 1$ and $0  \le   v_\mathrm{m} < 1$. Thus, the gap depends only on the modulation velocity $v_\mathrm{m}$ and the vector-to-scalar offset ratio $r_{\ratioAV}$: $\Delta_V = 2m$ at $v_\mathrm{m} = 0$ (static limit), while $\Delta_V \to 0$ as $v_\mathrm{m} \to 1^{-}$ (closure limit). By controlling $v_\mathrm{m}$ and $r_{\ratioAV}$, the gap is continuously tunable. At fixed $v_\mathrm{m}$, $\Delta_V$ grows monotonically with $r_{\ratioAV}$ and, as $r_{\ratioAV} \to 1^{-}$, approaches $\Delta_V^{\max}(v_\mathrm{m}) = 2m\,e^{\omega_{\mathrm m}}$ with the modulation rapidity $\omega_{\mathrm m} \equiv \operatorname{arctanh}(v_\mathrm{m})$. At fixed $r_{\ratioAV}$, the maximum occurs at $v_\mathrm{m} = r_{\ratioAV}$ with $\Delta_V^{\max}(r_{\ratioAV}) = 2m/\sqrt{1-r_{\ratioAV}^{2}}$. Hence, the gap collapses to zero with $v_\mathrm{m} \to 1^{-}$ at any fixed $r_{\ratioAV}<1$, or can be made arbitrarily large with $v_\mathrm{m} \to 1^{-}$ and $r_{\ratioAV} \to 1^{-}$. The same control may extend to platforms where fast gating emulates space-time modulation, such as graphene and other Dirac materials, and to quantum simulators with programmable drives; in these settings, an effective modulation velocity $v_\mathrm{m}$ and the vector-to-scalar offset ratio $r_{\ratioAV}$ could be varied to continuously tune the interbranch evanescent window.

\pati{Oblique-Incidence Outlook} 
The $(1+1)$-dimensional model assumed in this paper is inherently restricted to normal incidence. Extending the analysis to $(2+1)$ dimensions would introduce a conserved transverse momentum and render the threshold and transmission angle-dependent. This would allow investigation of whether space--time modulation can support super-Klein tunneling, namely omnidirectional perfect transmission. In pseudospin-1 systems, such all-angle transmission is enabled by special matching conditions associated with the three-component pseudospin structure ~\cite{Urban_2011_Barrier} and has been observed experimentally in phononic crystals~\cite{Zhu_2023_Experimental}. The present model, by contrast, has a two-component Dirac spinor, so establishing an analogous all-angle transmission condition would require a dedicated $(2+1)$-dimensional analysis.

\section{Velocity-dependent Klein paradox}
\pati{Velocity-dependent Klein Paradox}
At fixed step height, scanning the modulation velocity $v_\mathrm{m}$ [vertical axis of Fig.~\ref{fig:ST_probability}(a)] reveals a nonmonotonic transmission: it vanishes over a finite window $\Delta_{v_\mathrm{m}}$---an evanescent interval that persists even when the electron can catch up with the modulation ($v_\mathrm{m} < v_\mathrm{g}$)---then reopens into the negative-energy continuum and increases as $v_\mathrm{m} \to 1^{-}$. We may term this on/off/on behavior a \emph{velocity-dependent Klein paradox}, distinct from the kinematic no-catch-up band ($v_\mathrm{m} > v_\mathrm{g}$). This behavior is akin in nature and significance to the conventional Klein paradox.

\section{Conclusion}
\pati{Conclusion}
We introduce space-time modulation of the four-potential as a scattering knob that enforces oblique energy-momentum transitions at a modulation-step interface. Within a subluminal framework, we derive closed-form lab-frame kinematics for the outgoing channels, determine reflection and transmission from spinor continuity and compute measurable probabilities by projecting the Dirac current onto the modulation interface. The analysis reveals a drastic reduction in the potential required to access Klein tunneling compared with the static-step threshold that demands continuum overlap, establishes that the Klein gap is fully engineerable via the modulation velocity and the vector-to-scalar potential offset ratio, and uncovers a velocity-dependent Klein paradox in which transmission vanishes within a finite velocity window and reemerges, increasing as the front approaches light speed. These results indicate experimentally attainable conditions for accessing Klein tunneling using space-time programmable fronts, opening avenues for tunable electron-wave manipulation.

\section*{Acknowledgements}
This work is supported by FWO under Grant G0B0623N.

\bibliographystyle{elsarticle-num} 
\bibliography{Klein_Paradox}






\makeatletter
\newcommand{\supplementarycontents}{%
  \section*{Contents}%
  \@starttoc{stc}%
}
\newcommand{\suppsection}[1]{%
  \section{#1}%
  \addcontentsline{stc}{section}{\protect\numberline{\thesection}#1}%
}
\newcommand{\suppsubsection}[1]{%
  \subsection{#1}%
  \addcontentsline{stc}{subsection}{\protect\numberline{\thesubsection}#1}%
}
\newcommand{\suppsubsubsection}[1]{%
  \subsubsection{#1}%
  \addcontentsline{stc}{subsubsection}{\protect\numberline{\thesubsubsection}#1}%
}

\makeatother

\clearpage

\setcounter{secnumdepth}{3}
\setcounter{section}{0}
\setcounter{subsection}{0}
\setcounter{subsubsection}{0}
\setcounter{equation}{0}
\setcounter{figure}{0}
\setcounter{table}{0}

\renewcommand{\thesection}{S\arabic{section}}
\renewcommand{\thesubsection}{\thesection.\arabic{subsection}}
\renewcommand{\thesubsubsection}{\thesubsection.\arabic{subsubsection}}

\makeatletter
\renewcommand{\p@section}{}%
\renewcommand{\p@subsection}{}%
\renewcommand{\p@subsubsection}{}%
\def\@seccntformat#1{\csname the#1\endcsname\quad}
\makeatother

\begin{center}
\textbf{Supplemental Material for ``Access to Klein Tunneling via Space-Time Modulation''}
\end{center}
\vspace{0.5em}

\setcounter{tocdepth}{3}
\supplementarycontents
\clearpage

\suppsection{Dirac Equation}\label{sec:s_DE}

\pati{Derivation of 1+1D Dirac Equation}

We use the metric $\eta^{\mu\nu}=\mathrm{diag}(1,-1,-1,-1)$. The minimally coupled Dirac equation is~\cite{Dirac_1928,Greiner_1985_QED}
\begin{equation}\label{eq:s_Dirac4D}
    \left[\gamma^\mu\left(i\partial_\mu - q A_\mu\right)-m\right]\psi=0,
\end{equation}
with $\{\gamma^\mu,\gamma^\nu\}=2\,\eta^{\mu\nu}$, where $A^\mu=(V,\mathbf A)$ so $A_\mu=(V,-\mathbf A)$. Multiplying \eqref{eq:s_Dirac4D} on the left by $\gamma^0$ yields the Hamiltonian form
\begin{equation}
    i\partial_t\psi
    =\left[-i\,\gamma^0\gamma^i\partial_i + q A_0 + q\,\gamma^0\gamma^i A_i + \gamma^0 m\right]\psi
    =\left[\alpha^i(-i\partial_i + q A_i) + \beta\,m + qV\right]\psi,
\label{S:H_general}
\end{equation}
where $\beta\equiv\gamma^0$, $\alpha^i\equiv\gamma^0\gamma^i$ $(i=1,2,3)$, $A_0=V$, and $A_i=-A^i$.
Equivalently, in 3-vector notation,
\begin{equation}
i\partial_t\psi=\left[\boldsymbol{\alpha}\!\cdot\!\left(-i\nabla - q\,\mathbf A\right)+\beta\,m+qV\right]\psi.
\end{equation}

For dynamics restricted to $z$, with a longitudinal vector potential,
\begin{equation}
\partial_x=\partial_y=0,\qquad A_x=A_y=0,\qquad A_z\equiv A(t,z),
\end{equation}
and \eqref{S:H_general} reduces to
\begin{equation}
i\partial_t\psi=\left[\alpha^{3}\left(-i\partial_z - qA\right)+\beta\,m+qV\right]\psi,
\label{S:H_1D_block}
\end{equation}
where, in the Dirac-Pauli representation,
\begin{equation}
\gamma^{0}=\begin{pmatrix} I&0\\0&-I\end{pmatrix}~\text{and}~\gamma^{3}=\begin{pmatrix} 0&\sigma^{3}\\-\sigma^{3}&0\end{pmatrix}
\ \text{imply}~
\alpha^{3}=\gamma^{0}\gamma^{3}=
\begin{pmatrix} 0&\sigma^{3}\\\sigma^{3}&0\end{pmatrix}.
\end{equation}

This decouples the $4\times4$ system~\eqref{S:H_1D_block} into the identical equation sets that reduce to the $2\!\times\!2$-block system
\begin{subequations}\label{eq:Dirac_Sch_form}
\begin{equation}
i\partial_t\psi=H\psi,
\end{equation}
where
\begin{equation}\label{S:H_2x2}
H=
\begin{pmatrix}
m+qV & -i\partial_z - qA\\
-i\partial_z - qA & -m+qV
\end{pmatrix}
= \sigma_x\left(-i\,\partial_z - qA\right)+\sigma_z\,m+qV,
\end{equation}
\end{subequations}
with the Pauli matrices
\begin{equation}\label{eq:Pauli_matrix}
        \sigma_x=
        \begin{pmatrix}
            0 & 1 \\
            1 & 0 
        \end{pmatrix},
        \quad
        \sigma_y=
        \begin{pmatrix}
            0 & -i \\
            i & 0 
        \end{pmatrix}
        \quad\text{and}\quad
        \sigma_z=
        \begin{pmatrix}
            1 & 0 \\
            0 & -1 
        \end{pmatrix}.
    \end{equation} 

\pati{Solution of 1+1D Dirac Equation}

We now assume a region where $V$ and $A$ are constant. Inserting a monochromatic plane-wave ansatz
\begin{equation}\label{S:ansatz}
\psi(t,z)=u(p)\,e^{-i(Et-pz)},~\text{with}~
u(p)=\begin{pmatrix}u_1\\ u_2\end{pmatrix},
\end{equation}
with the substitutions $i\partial_t\!\to\!E$ and $-i\partial_z\!\to\!p$ into Eq.~\eqref{eq:Dirac_Sch_form} yields the eigen-equation
\begin{subequations}
\begin{equation}\label{eq:s_Dirac2D_matrix}
E\,u(p)=
\begin{pmatrix}
m+qV & p-qA\\
p-qA & -m+qV
\end{pmatrix}u(p),
\end{equation}
or, equivalently,
\begin{equation}\label{S:Pauli}
\left[\sigma_x\,(p-qA)+\sigma_z\,m\right]\,u=(E-qV)\,u,
\end{equation}
or, in components,
\begin{equation}\label{S:comp}
(m-(E-qV))u_1+(p-qA) u_2=0~\text{and}~
(p-qA) u_1-(m+(E-qV))u_2=0.
\end{equation}
\end{subequations}

A nontrivial solution of this equation requires
\begin{subequations}
\begin{equation}\label{S:disp}
(E-qV)^2=(p-qA)^2+m^2
\end{equation}
and
\begin{equation}\label{S:ratio}
\frac{u_2}{u_1}=\frac{(E-qV)-m}{(p-qA)}
=\frac{(p-qA)}{(E-qV)+m}.
\end{equation}
\end{subequations}

In the (unnormalized) basis with $u_1=1$, an eigen-spinor is found as
\begin{subequations}
\begin{equation}
u(p)=
\begin{pmatrix}
1\\\dfrac{E-qV-m}{\,p-qA\,}
\end{pmatrix},
\end{equation}
which leads, using Eq.~\eqref{S:ansatz}, to
\begin{equation}\label{S:psi_pm_final}
\psi(t,z)=
\begin{pmatrix}
1\\\dfrac{E-qV-m}{\,p-qA\,}
\end{pmatrix} e^{-i(Et-pz)}.
\end{equation}
\end{subequations}

\clearpage

\suppsection{Subluminal Energy-Momentum Transitions}\label{sec:s_sub_EP}

\pati{Energy-Momentum in Terms of Inputs}

The momenta of the incident, reflected, and transmitted electron are found from the dispersion relation~\eqref{S:disp} as
\begin{subequations}\label{eq:s_p_disper}
\begin{equation}\label{eq:s_pi_disper}
    p_\mathrm{i}=\sqrt{(E_\mathrm{i}-qV_1)^2 - m^2} + qA_1,
\end{equation}
\begin{equation}\label{eq:s_pr_disper}
    p_\mathrm{r}=\pm\sqrt{(E_\mathrm{r}-qV_1)^2 - m^2} + qA_1,
\end{equation}
\begin{equation}\label{eq:s_pt_disper}
    p_\mathrm{t}=\pm\sqrt{(E_\mathrm{t}-qV_2)^2 - m^2} + qA_2.
\end{equation}
\end{subequations}
According to Noether's theorem~\cite{Noether_1918}, in the comoving frame, energy is conserved ($\Delta E^{\prime}=0$) due to temporal translational symmetry, i.e.,
\begin{equation}\label{eq:s_Eprime_cons}
    E^{\prime}_\mathrm{i}=E^{\prime}_\mathrm{r}=E^{\prime}_\mathrm{t}.
\end{equation}
Applying Lorentz transformations~\cite{Greiner_1985_QED} to these relations leads to the lab frame relations
\begin{equation}\label{eq:s_E_Lorentz}
    \gamma_\mathrm{m}(E_\mathrm{i} - v_\mathrm{m}p_\mathrm{i}) = \gamma_\mathrm{m}(E_\mathrm{r} - v_\mathrm{m}p_\mathrm{r}) = \gamma_\mathrm{m}(E_\mathrm{t} - v_\mathrm{m}p_\mathrm{t}),
\end{equation}
which, upon insertion of Eqs.~\eqref{eq:s_p_disper} into Eq.~\eqref{eq:s_E_Lorentz}, yield the two equations
\begin{equation}\label{eq:s_Er_inter}
    E_\mathrm{r} - v_\mathrm{m}\left(\pm\sqrt{(E_\mathrm{r}-qV_1)^2 - m^2} + qA_1\right) = E_\mathrm{i} - v_\mathrm{m}p_\mathrm{i},
\end{equation}
and
\begin{equation}\label{eq:s_Et_inter}
    E_\mathrm{t} - v_\mathrm{m}\left(\pm\sqrt{(E_\mathrm{t}-qV_2)^2 - m^2} + qA_2\right) = E_\mathrm{i} - v_\mathrm{m}p_\mathrm{i}.
\end{equation}
Solving Eq.~\eqref{eq:s_Er_inter} for $E_\mathrm{r}$ yields
\begin{equation}\label{eq:s_Er}
    E_\mathrm{r} = E_\mathrm{i} + 2\gamma_\mathrm{m}^2 \left[v_\mathrm{m}^2 (E_\mathrm{i}-qV_1) - v_\mathrm{m} (p_\mathrm{i}-qA_1)\right],
\end{equation}
while solving Eq.~\eqref{eq:s_Et_inter} for $E_\mathrm{t}$ yields
\begin{equation}\label{eq:s_Et}
    E_\mathrm{t}^{\pm} = E_\mathrm{i} + \gamma_\mathrm{m}^2 \left [v_\mathrm{m}^2 (E_\mathrm{i}-qV_2) - v_\mathrm{m}\left(\left(p_\mathrm{i} - qA_2\right) \mp \sqrt{\left(E_\mathrm{i}-qV_2 - v_\mathrm{m} \left(p_\mathrm{i} - qA_2\right)\right)^2 - (m/\gamma_\mathrm{m})^2}\right)\right].
\end{equation}
Inserting Eq.~\eqref{eq:s_Er} and Eq.~\eqref{eq:s_Et} into Eq.~\eqref{eq:s_E_Lorentz} separately provides the momenta
\begin{equation}
    p_\mathrm{r}=p_\mathrm{i} + 2\gamma_\mathrm{m}^2 \left[v_\mathrm{m}(E_\mathrm{i}-qV_1)-(p_\mathrm{i}-qA_1) \right],
\end{equation}
and
\begin{equation} 
    p_\mathrm{t}^{\pm}=  \gamma_\mathrm{m}^2 v_\mathrm{m} \left[\left(E_\mathrm{i}-qV_2 \right) - v_\mathrm{m} \left(p_\mathrm{i} - qA_2\right)\right] \pm \gamma_\mathrm{m}^2 \sqrt{\left[\left(E_\mathrm{i}-qV_2 \right) - v_\mathrm{m} \left(p_\mathrm{i} - qA_2\right)\right]^2 - (m/\gamma_\mathrm{m})^2} + qA_2,
\end{equation}
where
\begin{equation}
    \gamma_\mathrm{m}=1/\sqrt{1-v_\mathrm{m}^2}.
\end{equation}

These results can be written in the more compact form
\begin{subequations}\label{eq:s_lab_var}
    \begin{equation}
    E_\mathrm{r} = E_\mathrm{i} + 2\gamma_\mathrm{m}^{2} v_\mathrm{m}  W_\mathrm{r},
    \end{equation}
    \begin{equation}
    p_\mathrm{r}=p_\mathrm{i} + 2\gamma_\mathrm{m}^{2} W_\mathrm{r},
    \end{equation}
    \text{where} $W_\mathrm{r} = v_\mathrm{m} (E_\mathrm{i}-qV_1) - (p_\mathrm{i}-qA_1)$, \text{and}
    \begin{equation}
    E_\mathrm{t}^{\pm} = \gamma_\mathrm{m}^{2} W_\mathrm{t} \pm \gamma_\mathrm{m}^{2} v_\mathrm{m} \sqrt{W_\mathrm{t}^2 - \left(m/\gamma_\mathrm{m}\right)^2} + qV_2,
    \end{equation}
    \begin{equation}\label{eq:s_pt}
    p_\mathrm{t}^{\pm} = \gamma_\mathrm{m}^{2} v_\mathrm{m} W_\mathrm{t} \pm \gamma_\mathrm{m}^{2} \sqrt{W_\mathrm{t}^2 - \left(m/\gamma_\mathrm{m}\right)^2}  + qA_2,
\end{equation}
\end{subequations}
where $W_\mathrm{t} = (E_\mathrm{i}-qV_2) - v_\mathrm{m} \left(p_\mathrm{i} - qA_2\right)$.

\clearpage

\suppsection{Signs of the Energies and Momenta}\label{sec:s_sub_sign}
\pati{Definition of the Critical Velocities}

\begin{figure}[ht!]
    \centering
    \includegraphics[width=0.65\textwidth]{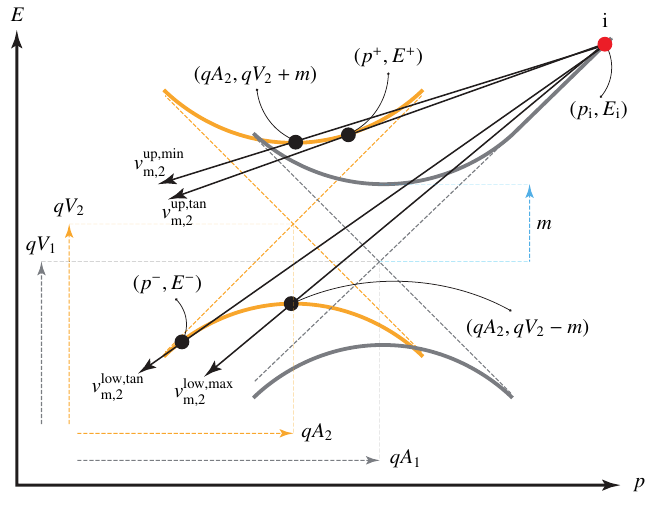}
    \vspace{-4mm}
    \caption{Critical modulation velocities. Dispersions $E(p)$ of medium~1 (gray) and medium~2 (brown). The transition lines from the incident state (red) define the critical slopes, equal to the modulation velocities, $v_\mathrm{m,2}^{\mathrm{up,min}}$, $v_\mathrm{m,2}^{\mathrm{low,max}}$, $v_\mathrm{m,2}^{\mathrm{up,tan}}$ and $v_\mathrm{m,2}^{\mathrm{low,tan}}$.}\label{fig:s_critical_vel}
\end{figure}

Figure~\ref{fig:s_critical_vel} shows the superimposed dispersion diagrams of the two media---two hyperbolas given by Eq.~\eqref{S:disp}, the first-medium one in gray and the second-medium one in brown. From the incident state, represented by the red point, we have drawn straight ``transition lines'' to specific points on the second dispersion curve. The slopes of these lines define the critical modulation velocities and, in turn, determine the appropriate signs in Eqs.~\eqref{eq:s_lab_var}. We identify the critical velocities by the intersection points of the transition line as follows:
\begin{itemize}
    \item the lowest-energy point of the upper dispersion branch of the second medium, $v_\mathrm{m,2}^{\mathrm{up,min}}$;
    \item the highest-energy point of the lower dispersion branch of the second medium, $v_\mathrm{m,2}^{\mathrm{low,max}}$;
    \item the tangent point of the upper dispersion branch of the second medium, $v_\mathrm{m,2}^{\mathrm{up,tan}}$;
    \item the tangent point of the lower dispersion branch of the second medium, $v_\mathrm{m,2}^{\mathrm{low,tan}}$.
\end{itemize}

\suppsubsection{$v_\mathrm{m,2}^{\mathrm{up,min}}$ and $v_\mathrm{m,2}^{\mathrm{low,max}}$}
\pati{Highest-lowest Energy Modulation Velocities}

From the geometry in Fig.~\ref{fig:s_critical_vel}, each critical velocity is the slope
$\Delta E/\Delta p$ of the transition line from the incident point to the corresponding indicated
intersection. Hence
\begin{equation}
    v_\mathrm{m,2}^\mathrm{up,min}=\frac{E_\mathrm{i}-qV_2-m}{p_\mathrm{i}-qA_2}
    \qquad\text{and}\qquad
    v_\mathrm{m,2}^\mathrm{low,max}=\frac{E_\mathrm{i}-qV_2+m}{p_\mathrm{i}-qA_2}.
\end{equation}
These expressions follow by reading off the vertical and horizontal separations
$(\Delta E,\Delta p)$ between the incident state and the corresponding intersection points.

\suppsubsection{$v_\mathrm{m,2}^{\mathrm{up,tan}}$ and $v_\mathrm{m,2}^{\mathrm{low,tan}}$}
\pati{Tangent Modulation Velocities}
We can find the velocities $v_\mathrm{m,2}^{\mathrm{up,tan}}$ and $v_\mathrm{m,2}^{\mathrm{low,tan}}$ in two ways: a geometrical way and a momentum evanescence way.

\begin{figure}[h!]
    \centering
    \includegraphics[width=0.65\textwidth]{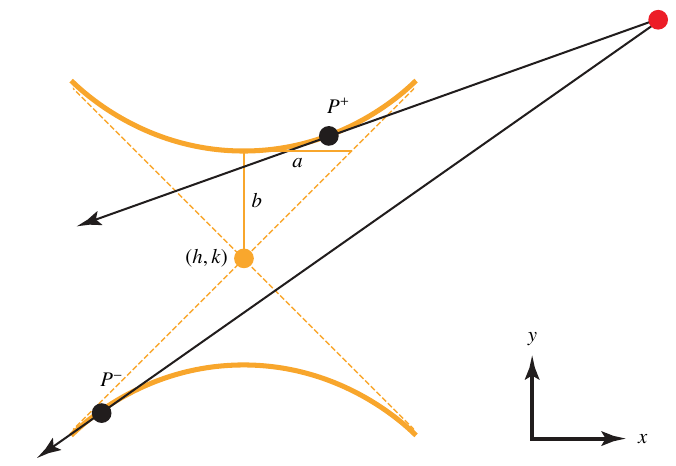}
    \caption{Medium-2 hyperbola and tangent lines from the point i in Fig.~\ref{fig:s_critical_vel}, with tangent points $P^+$ and $P^-$.}\label{fig:s_hyperbola}
\end{figure}

\suppsubsubsection{Calculation using Geometry}
\pati{First Method for Tangent Modulation Velocities}
Figure~\ref{fig:s_hyperbola} shows the medium-2 hyperbola and tangent lines from the point i in Fig.~\ref{fig:s_critical_vel}, where the two tangency points are denoted $P^+$ and $P^-$. We shall determine the coordinates of these points and the slopes at these points, corresponding to the sought-after quantities $v_\mathrm{m,2}^{\mathrm{up,tan}}$ and $v_\mathrm{m,2}^{\mathrm{low,tan}}$.

The equation of the hyperbola and straight lines in Fig.~\ref{fig:s_hyperbola} are
\begin{equation}\label{eq:s_hyper_eq}
    \frac{(y - k)^2}{a^2} - \frac{(x - h)^2}{b^2} = 1,
\end{equation}
and
\begin{equation}\label{eq:s_line_eq}
    y = nx + c,
\end{equation}
respectively. Substituting Eq.~\eqref{eq:s_line_eq} into \eqref{eq:s_hyper_eq} yields the intersection equation
\begin{equation}
    \frac{(nx + c - k)^2}{a^2} - \frac{(x - h)^2}{b^2} = 1,
\end{equation}
whose expansion yields the quadratic equation
\begin{subequations}\label{eq:s_hyper_line_exp}
\begin{equation}
    Ax^2 + Bx + C = 0,
\end{equation}
where
\begin{equation}\label{eq:ABC}
    A = (b^2n^2 - a^2), \quad B = \left[ 2b^2n(c - k) + 2a^2h \right] \quad\text{and} \quad C = b^2(c - k)^2 - a^2(h^2 + b^2).
\end{equation}
\end{subequations}
We find the points $P^\pm$---corresponding to a single intersection point for each straight line---by nullifying the discriminant of Eq.~\eqref{eq:s_hyper_line_exp}, viz.,
\begin{equation}
    \Delta = B^2 - 4AC = 0,
\end{equation}
i.e., using Eq.~\eqref{eq:ABC},
\begin{equation}
    \left[2b^2n(c - k) + 2a^2h \right]^2 - 4(b^2n^2 - a^2) \left[ b^2(c - k)^2 - a^2(h^2 + b^2) \right] = 0.
\end{equation}
Solving for the $y$-intercept $c$ then gives the new quadratic equation
\begin{subequations}
\begin{equation}
    Dc^2 + Ec + F = 0,
\end{equation}
where
\begin{equation}
    D = 4a^2b^2n^2, \quad E = \left[8a^2b^2n(h - k)\right] \quad\text{and}\quad F = 4a^2b^2\left[-a^2 + b^2n^2 + h^2n^2 - 2hkn + k^2\right],
\end{equation}
\end{subequations}
whose solutions are
\begin{equation}
    c^{\pm} = \frac{-E \pm \sqrt{E^2 - 4DF}}{2D},
\end{equation}
i.e., explicitly,
\begin{equation}
    c^{\pm} = -hn + k \pm \sqrt{a^2 - b^2n^2}.
\end{equation}
Inserting $c^{\pm}$ into Eq.~\eqref{eq:s_line_eq} yields the two tangent lines
\begin{equation}\label{eq:s_y_1,2}
    y^{\pm} = nx - hn + k \pm \sqrt{a^2 - b^2n^2},
\end{equation}
whose substitution into Eq.~\eqref{eq:s_hyper_eq} gives the $x$-coordinates of the tangency points,
\begin{subequations}\label{eq:P_coord}
\begin{equation}
    x^{\pm} = h \pm \frac{b^2 n}{\sqrt{a^2 - b^2 n^2}},
\end{equation}
whose substitution into \eqref{eq:s_y_1,2} gives the corresponding $y$-coordinates,
\begin{equation}
    y^{\pm} = k \pm \frac{a^2}{\sqrt{a^2 - b^2 n^2}},
\end{equation}
\end{subequations}
where one notes that real tangency points require $a^2 - b^2 n^2>0$.

Comparing the hyperbola equation~\eqref{eq:s_hyper_eq} with the dispersion relation~\eqref{S:disp} reveals the correspondence
\begin{subequations}
\begin{equation}
    y \rightarrow E, \quad x \rightarrow p, \quad k \rightarrow qV, \quad h \rightarrow qA, \quad a \rightarrow m, \quad b \rightarrow m,
\end{equation}
and, knowing that the slope of the dispersion curves is $v_\mathrm{m}$ indicates that the slope of the tangent straight lines in Eq.~\eqref{eq:s_line_eq} is
\begin{equation}
    n \rightarrow v_\mathrm{m}.
\end{equation}
\end{subequations}
Under this mapping, the coordinates of the tangent points in the energy-momentum dispersion diagram are found to be
\begin{subequations}\label{eq:s_tan_points}
\begin{equation}
    (p^+, E^+) = \left(qA_2 + m v_\mathrm{m}^+ \gamma_\mathrm{m}^+,\, qV_2 + m \gamma_\mathrm{m}^+\right),
\end{equation}
\begin{equation}
    (p^-, E^-) = \left(qA_2 - m v_\mathrm{m}^- \gamma_\mathrm{m}^-,\, qV_2 - m \gamma_\mathrm{m}^-\right),
\end{equation}
\end{subequations}
where $\gamma_\mathrm{m}=1/\sqrt{1 - v_\mathrm{m}^2}$.

The slopes at these points then yield, by geometry (Fig.~\ref{fig:s_critical_vel}), the critical tangent modulation velocities
\begin{subequations}\label{eq:s_critical_vel_inter}
\begin{equation}
    v_\mathrm{m}^+ = \frac{E_\mathrm{i} - E^+}{p_\mathrm{i} - p^+},
\end{equation}
\begin{equation}
    v_\mathrm{m}^- = \frac{E_\mathrm{i} - E^-}{p_\mathrm{i} - p^-},
\end{equation}
\end{subequations}
which become upon substituting Eqs.~\eqref{eq:s_tan_points}
\begin{equation}\label{eq:s_gap_edges_vel}
    v_\mathrm{m}^{\pm} = \frac{E_\mathrm{i} - qV_2 \mp m\gamma_\mathrm{m}^{\pm}}{p_\mathrm{i} - qA_2 \mp m v_\mathrm{m}^{\pm}\gamma_\mathrm{m}^{\pm}}.
\end{equation}
Finally, solving this equation for $v_\mathrm{m}^{\pm}$ yields
\begin{equation}\label{eq:s_v_m}
    v_\mathrm{m}^{\pm} = \frac{(p_\mathrm{i} - qA_2)(E_\mathrm{i} - qV_2) \mp m\sqrt{(p_\mathrm{i} - qA_2)^2 - (E_\mathrm{i} - qV_2)^2 + m^2}}{(p_\mathrm{i} - qA_2)^2 + m^2},
\end{equation}
where $v_\mathrm{m}^{+}$ is the tangent point of the upper dispersion branch, $v_\mathrm{m,2}^{\mathrm{up,tan}}$, and $v_\mathrm{m}^{-}$ is the tangent point of the lower dispersion branch, $v_\mathrm{m,2}^{\mathrm{low,tan}}$. 

\suppsubsubsection{Calculation using the Evanescent Transmitted Momentum}
\pati{Second Method for Tangent Modulation Velocities}

Increasing $v_\mathrm{m}$ in the equation of the momentum, given by Eq.~\eqref{eq:s_pt}, corresponds to changing the slope of the transition lines from the i point in Fig.~\ref{fig:s_critical_vel}, from a horizontal line ($v_\mathrm{m}=0$) to a vertical line ($v_\mathrm{m}=\infty$). When this line intersects the gap, then the momentum becomes complex, and the limit velocity where this occurs corresponds to the tangencies illustrated in Fig.~\ref{fig:s_hyperbola}. Therefore, we may obtain the tangency points by setting the square root in Eq.~\eqref{eq:s_pt} to zero, viz.,
\begin{equation}
    \left[\left(E_\mathrm{i}-qV_2 \right) - v_\mathrm{m} \left(p_\mathrm{i} - qA_2\right)\right]^2 - (m/\gamma_\mathrm{m})^2 = 0.
\end{equation}
Solving this equation for $v_\mathrm{m}$, while holding the other variables fixed, reproduces Eq.~\eqref{eq:s_v_m}.

\suppsubsection{Choosing the Signs of the Energies and Momenta}
\pati{Choosing the Signs of the Energies and Momenta}

The signs to be chosen in Eq.~\eqref{eq:s_lab_var} for physicality depend on $v_\mathrm{m}$ (slope) intervals as follows, considering that the positive (resp. negative) superscripts in the pairs $(E_\mathrm{t}^\pm,p_\mathrm{t}^\pm)$ correspond to the right (resp. left) of the medium~2 hyperbola in Fig.~\ref{fig:s_critical_vel}. 
\begin{itemize}
    \item $0 < v_\mathrm{m} < v_{\mathrm{m},2}^{\mathrm{up,min}}$:\; Only the pair $(E_\mathrm{t}^{+},p_\mathrm{t}^{+})$ has the required positive group velocity for the transmitted wave, $v_\mathrm{g,t}=\pdv{E_\mathrm{t}}{p_\mathrm{t}}=\frac{p_\mathrm{t}-qA_2}{E_\mathrm{t}-qV_2}$. 

    \item $v_{\mathrm{m},2}^{\mathrm{up,min}} < v_\mathrm{m} < v_{\mathrm{m},2}^{\mathrm{up,tan}}$:\; Both pairs $(E_\mathrm{t}^\pm,p_\mathrm{t}^\pm)$ have a positive group velocity. However, the pair $(E_\mathrm{t}^{-},p_\mathrm{t}^{-})$ is associated with $v_{\mathrm{g,t}} < v_\mathrm{m}$, which prevents the existence of a transmitted wave. In contrast, the pair $(E_\mathrm{t}^{+},p_\mathrm{t}^{+})$ satisfies the transmission requirement $v_{\mathrm{g,t}} > v_\mathrm{m}$.

    \item $v_{\mathrm{m},2}^{\mathrm{up,tan}} < v_\mathrm{m} < v_{\mathrm{m},2}^{\mathrm{low,tan}}$:\; No transmission (Klein gap) occurs in this range.

    \item $v_{\mathrm{m},2}^{\mathrm{low,tan}} < v_\mathrm{m} < v_{\mathrm{m},2}^{\mathrm{low,max}}$:\; Both pairs $(E_\mathrm{t}^\pm,p_\mathrm{t}^\pm)$ have a positive group velocity. However, the pair $(E_\mathrm{t}^{+},p_\mathrm{t}^{+})$ is associated with $v_{\mathrm{g,t}} < v_\mathrm{m}$, which prevents the existence of a transmitted wave. In contrast, the pair $(E_\mathrm{t}^{-},p_\mathrm{t}^{-})$ satisfies the transmission requirement $v_{\mathrm{g,t}} > v_\mathrm{m}$.

    \item $v_{\mathrm{m},2}^{\mathrm{low,max}} < v_\mathrm{m} < v_{\mathrm{g}}$:\; Only the pair $(E_\mathrm{t}^{-},p_\mathrm{t}^{-})$ has the required positive group velocity for the transmitted wave; for reference, the incident-wave group velocity is $v_\mathrm{g}=\pdv{E_\mathrm{i}}{p_\mathrm{i}}=\frac{p_\mathrm{i}-qA_1}{E_\mathrm{i}-qV_1}$.
\end{itemize}

\clearpage

\suppsection{Relativistic Frame Transformations}\label{sec:s_LT}

\pati{Lorentz and spinor boosts used in the main text}

For the uniformly moving subluminal interface considered in the main text, we use the standard Lorentz boost along the $z$ axis. Writing the rapidity as $\omega$ and $\beta=v/c$, the coordinate transformation is
\begin{equation}
    t^{\prime} = \gamma \left( t - \frac{v}{c^2} z \right),
    \qquad
    z^{\prime} = \gamma (z - v t),
\end{equation}
with $\gamma=(1-\beta^2)^{-1/2}$. The corresponding Dirac-spinor transformation is
\begin{equation}
    \psi^{\prime}(x^{\prime}) = S_z(\Lambda)\,\psi(x),
\end{equation}
where, in the Dirac-Pauli representation,
\begin{equation}\label{eq:s_spinor_trans}
    S_z(\Lambda)=\cosh\!\left(\frac{\omega}{2}\right)-\alpha^3\sinh\!\left(\frac{\omega}{2}\right)
    =
        \begin{pmatrix}
            c_1 & 0 & -c_2 & 0 \\
            0 & c_1 & 0 & c_2 \\
            -c_2 & 0 & c_1 & 0 \\
            0 & c_2 & 0 & c_1
        \end{pmatrix},
\end{equation}
with $c_1=\sqrt{(\gamma+1)/2}$ and $c_2=\sqrt{(\gamma-1)/2}$. These are the standard boost relations used to pass between the laboratory and comoving descriptions in the main text~\cite{Greiner_1985_QED}.

\clearpage

\suppsection{Scattering Coefficients}

\suppsubsection{Spinor Continuity and Amplitudes}\label{sec:s_sub_amplit}
\pati{Spinor Continuity and Amplitudes}

The boundary condition---continuity of the spinor wave function, Eq.~\eqref{S:psi_pm_final}---is imposed in the $K^{\prime}$ frame. Accordingly, we express the spinor in $K^{\prime}$ and use the Lorentz spinor boost along $z$, Eq.~\eqref{eq:s_spinor_trans}. The spinor transforms from $K$ to $K^{\prime}$ as
\begin{equation}
    \psi^{\prime}(t^{\prime},z^{\prime}) = S_z(\Lambda) \psi(t,z),
\end{equation}
which reduces for a 1+1D plane-wave to
\begin{equation} \label{eq:s_psi_Kprime}
    \psi^{\prime}(t^{\prime},z^{\prime}) =
        \begin{pmatrix}
        c_1 - c_2\Gamma\\
        -c_2 + c_1\Gamma
        \end{pmatrix}
    \mathrm{e}^{-i(E^{\prime} t^{\prime} - p^{\prime} z^{\prime})},
\end{equation}
where $\Gamma=\frac{E-qV-m}{p-qA}$. On each side of the modulation, we evaluate $\Gamma$ with the corresponding energy-momentum and potentials, and write $\Gamma_\mathrm{i}$, $\Gamma_\mathrm{r}$, and $\Gamma_\mathrm{t}$ for the incident, reflected and transmitted branches, respectively.

For a step in the scalar and vector potentials, we set
\begin{equation} \label{eq:s_sub_V}
        (V,A) =
        \begin{cases}
        V_1, A_1 &\;\textrm{for}\;z^{\prime}<z^{\prime}_{0}, \\
        V_2, A_2 &\;\textrm{for}\;z^{\prime}>z^{\prime}_{0}.
    \end{cases}
\end{equation}
The left ($z^{\prime}<z^{\prime}_0$) spinor is
\begin{subequations}
\begin{equation}
    \psi^{\prime}_{1}(t^{\prime},z^{\prime}) = 
    \begin{pmatrix}
        c_1 - c_2\Gamma_\mathrm{i}\\
        -c_2 + c_1\Gamma_\mathrm{i}
    \end{pmatrix} \mathrm{e}^{-i E^{\prime}_\mathrm{i} t^{\prime}} \mathrm{e}^{i p^{\prime}_\mathrm{i} z^{\prime}}  
    +r \begin{pmatrix}
        c_1 - c_2\Gamma_\mathrm{r}\\
        -c_2 + c_1\Gamma_\mathrm{r}
    \end{pmatrix} \mathrm{e}^{-i E^{\prime}_\mathrm{r} t^{\prime}} \mathrm{e}^{i p^{\prime}_\mathrm{r} z^{\prime}},
\end{equation}
whose first and second terms correspond to the incident and reflected waves, with reflection amplitude $r$. The transmitted (right, $z^{\prime}>z^{\prime}_0$) spinor is
\begin{equation}
    \psi^{\prime}_{2}(t^{\prime},z^{\prime}) = 
    t \begin{pmatrix}
        c_1 - c_2\Gamma_\mathrm{t}\\
        -c_2 + c_1\Gamma_\mathrm{t}
    \end{pmatrix} \mathrm{e}^{-i E^{\prime}_\mathrm{t} t^{\prime}} \mathrm{e}^{i p^{\prime}_\mathrm{t} z^{\prime}},
\end{equation}
\end{subequations}
with transmission amplitude $t$.

In $K^{\prime}$, the modulation is static (pure-space) at $z^{\prime}=z^{\prime}_{0}$, so the spinor must be continuous there:
\begin{equation}
\left.\psi^{\prime}_{1}(t^{\prime},z^{\prime})\right|_{z^{\prime}=z^{\prime}_{0}}=\left.\psi^{\prime}_{2}(t^{\prime},z^{\prime})\right|_{z^{\prime}=z^{\prime}_{0}}.
\end{equation}
This yields the linear system
\begin{subequations}\label{eq:s_r_and_t_eqs}
    \begin{equation}
        1+r=t,
    \end{equation}
    \begin{equation}
        \Gamma_\mathrm{i}+r\Gamma_\mathrm{r}=t\Gamma_\mathrm{t}.
    \end{equation}
\end{subequations}
Solving Eqs.~\eqref{eq:s_r_and_t_eqs} gives
\begin{equation}
    r = \frac{\Gamma_\mathrm{i} - \Gamma_\mathrm{t}}{\Gamma_\mathrm{t} - \Gamma_\mathrm{r}},
\end{equation}
\begin{equation}
    t = \frac{\Gamma_\mathrm{i} - \Gamma_\mathrm{r}}{\Gamma_\mathrm{t} - \Gamma_\mathrm{r}}.
\end{equation}

\suppsubsection{Dirac Currents and Probabilities}\label{sec:s_sub_prob}
\pati{Dirac Currents and Probabilities}

The $1{+}1$D Dirac Hamiltonian $H=\sigma_x(-i\partial_z-qA)+\sigma_z m+qV$ (Eq.~\eqref{S:H_2x2}) yields the charge density and longitudinal current~\cite{Greiner_1985_QED}
\begin{equation}
    \rho=\psi^\dagger\psi,\qquad
    j^{\,z}=\psi^\dagger\sigma_x\psi.
\end{equation}
For a plane wave, $\psi=\begin{pmatrix}1\\ \Gamma\end{pmatrix}e^{-i(Et-pz)}$, the densities and currents are found to be
\begin{equation}
    \rho=\psi^\dagger \psi=1+|\Gamma|^2,\qquad
    j^{\,z}=\psi^\dagger\sigma_x \psi=\Gamma+\Gamma^{\,\ast}=2\Re\{\Gamma\}.
    \label{eq:rho-jz-Gamma}
\end{equation}
The flux through the space-time modulation worldline $f(t,z)=z-v_\mathrm{m}t=z_0$
uses the (unnormalized) covector $n_\mu\propto\partial_\mu f=(-v_\mathrm{m},1)$, so
\begin{equation}
    j_a = j_a^{\,\mu} n_\mu=j_a^{\,z}-v_\mathrm{m}\rho_a =2\Re\{\Gamma_a\}\;-\;v_\mathrm{m}\left(1+|\Gamma_a|^2\right),
    \label{eq:jn-Gamma}
\end{equation}
where $a={\mathrm i},{\mathrm r},{\mathrm t}$ labels the incident, reflected and transmitted channels, respectively.

With the spinor decomposition
\begin{equation}
    \psi_\mathrm{i}=\begin{pmatrix}1\\ \Gamma_\mathrm{i}\end{pmatrix}
    e^{-iE_\mathrm{i}t}e^{+ip_\mathrm{i}z},\qquad
    \psi_\mathrm{r}=r\,\begin{pmatrix}1\\ \Gamma_\mathrm{r}\end{pmatrix}
    e^{-iE_\mathrm{r}t}e^{+ip_\mathrm{r}z}\qquad
    \text{and}\qquad
    \psi_\mathrm{t}=t\,\begin{pmatrix}1\\ \Gamma_\mathrm{t}\end{pmatrix}
    e^{-iE_\mathrm{t}t}e^{+ip_\mathrm{t}z},
\end{equation}
the channel parameters are
\begin{equation}
    \Gamma_\mathrm{i}=\frac{E_\mathrm{i}-qV_1-m}{\,p_\mathrm{i}-qA_1\,},\quad
    \Gamma_\mathrm{r}=\frac{E_\mathrm{r}-qV_1-m}{\,p_\mathrm{r}-qA_1\,}\quad
    \text{and}\qquad
    \Gamma_\mathrm{t}=\frac{E_\mathrm{t}-qV_2-m}{\,p_\mathrm{t}-qA_2\,},
\end{equation}
with $(E_a-qV_\ell)^2=(p_a-qA_\ell)^2+m^2$ for the appropriate region $\ell=1,2$.

Upon averaging over rapid interference terms, the flux carried by each channel (scales $\propto$ amplitude$^2$) is
\begin{equation}
\begin{split}
    j_\mathrm{i}&
    =2\Re\{\Gamma_\mathrm{i}\}-v_\mathrm{m}\left(1+|\Gamma_\mathrm{i}|^2\right),\\
    j_\mathrm{r}&
    =|r|^2\left[2\Re\{\Gamma_\mathrm{r}\}-v_\mathrm{m}\left(1+|\Gamma_\mathrm{r}|^2\right)\right],\\
    j_\mathrm{t}&
    =|t|^2\left[2\Re\{\Gamma_\mathrm{t}\}-v_\mathrm{m}\left(1+|\Gamma_\mathrm{t}|^2\right)\right].
\end{split}
\end{equation}

The reflection and transmission probabilities are the flux ratios through the moving modulation,
\begin{equation}
    R=-\frac{j_\mathrm{r}}{j_\mathrm{i}}\qquad
    \text{and}\qquad
    T=\frac{j_\mathrm{t}}{j_\mathrm{i}},
    \label{eq:RT-def}
\end{equation}
and one verifies that $R+T=1$ using the amplitudes
$
r=\frac{\Gamma_\mathrm{i}-\Gamma_\mathrm{t}}{\Gamma_\mathrm{t}-\Gamma_\mathrm{r}}
~\text{and}~
t=\frac{\Gamma_\mathrm{i}-\Gamma_\mathrm{r}}{\Gamma_\mathrm{t}-\Gamma_\mathrm{r}}
$.
In the purely spatial limit $v_\mathrm{m}\to 0$, $j\to j^{\,z}$ and
Eq.~\eqref{eq:RT-def} reduces to the standard $R=-j^{\,z}_\mathrm{r}/j^{\,z}_\mathrm{i}$ and
$T=j^{\,z}_\mathrm{t}/j^{\,z}_\mathrm{i}$.

\clearpage

\suppsection{Threshold Potential}

\suppsubsection{Calculation of the Tangent Scalar Potential Offsets}\label{sec:s_sub_threshold}

\suppsubsubsection{Calculation using Modulation Velocity Fixing}
\pati{First Method for Calculation for Tangent Potentials}

Let us write Eq.~\eqref{eq:s_gap_edges_vel} in terms of the potential differences (offsets) $q\Delta V = qV_2-qV_1$ and $q\Delta A = qA_2-qA_1$ as
\begin{equation}\label{eq:s_v_m+-}
    v_\mathrm{m}^{\pm} = \frac{(E_\mathrm{i}-qV_1) - q\Delta V \mp m\gamma_\mathrm{m}^{\pm}}{(p_\mathrm{i}-qA_1) - q\Delta A \mp m v_\mathrm{m}^{\pm}\gamma_\mathrm{m}^{\pm}}.
\end{equation}

For a given energy $E_\mathrm{i}$ and modulation velocity $v_\mathrm{m}$, the Klein-gap region lies between two scalar-potential offsets $q\Delta V^{+}$ and $q\Delta V^{-}$. Assuming $q\Delta A^\pm = r_{\ratioAV}\, q\Delta V^\pm$, where $r_{\ratioAV}$ is real (and taken as fixed), equating the $\pm$ branches of Eq.~\eqref{eq:s_v_m+-} at the same $v_\mathrm{m}$ yields the two edge potentials
\begin{equation}\label{eq:s_Delta_V}
    q\Delta V^{\pm} = \frac{(E_\mathrm{i}-qV_1) - v_\mathrm{m} (p_\mathrm{i}-qA_1) \mp m/\gamma_\mathrm{m}}{1 - v_\mathrm{m} r_{\ratioAV}}.
\end{equation}
The edge associated with tangency to the lower-energy branch, $q\Delta V^{-}$, coincides with the Klein-tunneling threshold $q\Delta V^{\mathrm{th}}$.

\suppsubsubsection{Calculation using Evanescent Transmitted Momentum}
\pati{Second Method for Calculation for Tangent Potentials}

As an alternative calculation of the scalar potential offsets, we consider the imaginary limit of the square root term in Eq.~\eqref{eq:s_pt} that occurs between the two tangent values of the potential offset, since that interval corresponds to the evanescent channel. Consequently, the gap edges are obtained by setting the radicand to zero:
\begin{equation}
    \left[\left(E_\mathrm{i}-qV_2 \right) - v_\mathrm{m} \left(p_\mathrm{i} - qA_2\right)\right]^2 - (m/\gamma_\mathrm{m})^2 = 0.
\end{equation}
Solving this equation for $q\Delta V$, while holding the other variables fixed, reproduces Eq.~\eqref{eq:s_Delta_V}.

\suppsubsection{Velocity-Matching Threshold}\label{sec:s_sub_thresh_vel_match}
\pati{Velocity-Matching Threshold}

Using Eq.~\eqref{eq:s_Delta_V},
\begin{equation}\label{eq:s_Delta_V_th}
    q\,\Delta V^{\mathrm{th}} = \frac{(E_\mathrm{i}-qV_1) - v_\mathrm{m} (p_\mathrm{i}-qA_1) + m/\gamma_\mathrm{m}}{1 - v_\mathrm{m} r_{\ratioAV}},
\end{equation}
and the equations
\begin{subequations}
\begin{equation}\label{eq:s_Ei_gamma}
    E_\mathrm{i}-qV_1=\gamma_\mathrm{g}m
\end{equation}
and
\begin{equation}\label{eq:s_pi_gamma}
    p_\mathrm{i}-qA_1=\gamma_\mathrm{g} m v_\mathrm{g},
\end{equation}
\end{subequations}
and imposing the velocity-matching condition $v_\mathrm{m}=v_\mathrm{g}$ and $r_{\ratioAV}=-1$ yields
\begin{equation}
    q\,\Delta V^{\mathrm{th}} = \frac{\gamma_\mathrm{g}m - v_\mathrm{g} \gamma_\mathrm{g} m v_\mathrm{g} + m/\gamma_\mathrm{g}}{1 + v_\mathrm{g} }
    = 2m\sqrt{\frac{1-v_\mathrm{g}}{1+v_\mathrm{g}}}
    = 2m\,e^{-\omega_{\mathrm g}},
\end{equation}
with the electron rapidity $\omega_{\mathrm g}=\operatorname{arctanh}(v_{\mathrm g})$.

\suppsubsection{Velocity-Mismatch Sensitivity}
\label{sec:s_sub_thresh_mismatch}
\pati{Velocity-Mismatch Dependence}

\suppsubsubsection{Exact Threshold as a Function of Modulation Velocity}
\label{sec:s_sub_thresh_function_of_mod_vel}

We derive the threshold dependence away from exact velocity
matching. Starting from Eq.~\eqref{eq:s_Delta_V}, setting
$r_{\ratioAV}=-1$, and selecting the edge associated with tangency to the lower-energy branch gives
\begin{equation}
    q\,\Delta V^{\mathrm{th}}(v_\mathrm{m})
    =
    \frac{
    (E_\mathrm{i}-qV_1)
    -v_\mathrm{m}(p_\mathrm{i}-qA_1)
    +m/\gamma_\mathrm{m}
    }{
    1+v_\mathrm{m}
    }.
    \label{eq:s_thresh_vm_initial}
\end{equation}
Inserting Eqs.~\eqref{eq:s_Ei_gamma} and \eqref{eq:s_pi_gamma} into this equation yields
\begin{equation}
    q\,\Delta V^{\mathrm{th}}(v_\mathrm{m})
    =
    m
    \frac{
    \gamma_\mathrm{g}(1-v_\mathrm{m}v_\mathrm{g})
    +\gamma_\mathrm{m}^{-1}
    }{
    1+v_\mathrm{m}
    }.
    \label{eq:s_thresh_vm_velocity}
\end{equation}
Expressing then the electron and modulation velocities in terms of their rapidities,
\begin{equation}
    v_\mathrm{g}=\tanh\omega_\mathrm{g},
    \qquad
    v_\mathrm{m}=\tanh\omega_\mathrm{m},
\end{equation}
with
\begin{equation}
    \gamma_\mathrm{g}=\cosh\omega_\mathrm{g},
    \qquad
    \gamma_\mathrm{g}v_\mathrm{g}=\sinh\omega_\mathrm{g},
    \qquad
    \gamma_\mathrm{m}=\cosh\omega_\mathrm{m},
\end{equation}
transforms the first term of the numerator of
Eq.~\eqref{eq:s_thresh_vm_velocity} to
\begin{align}
    \gamma_\mathrm{g}(1-v_\mathrm{m}v_\mathrm{g})
    &=
    \cosh\omega_\mathrm{g}
    -
    \tanh\omega_\mathrm{m}\sinh\omega_\mathrm{g}
    \nonumber\\
    &=
    \frac{
    \cosh\omega_\mathrm{g}\cosh\omega_\mathrm{m}
    -
    \sinh\omega_\mathrm{g}\sinh\omega_\mathrm{m}
    }{
    \cosh\omega_\mathrm{m}
    }
    \nonumber\\
    &=
    \frac{
    \cosh(\omega_\mathrm{g}-\omega_\mathrm{m})
    }{
    \cosh\omega_\mathrm{m}
    }.
\end{align}
Moreover, the second term of the numerator and the denominator of Eq.~\eqref{eq:s_thresh_vm_velocity} can be written as
\begin{equation}
    \gamma_\mathrm{m}^{-1}
    =
    \frac{1}{\cosh\omega_\mathrm{m}},
    \qquad
    1+v_\mathrm{m}
    =
    \frac{e^{\omega_\mathrm{m}}}
    {\cosh\omega_\mathrm{m}}.
\end{equation}
Substituting these results into Eq.~\eqref{eq:s_thresh_vm_velocity} leads to
\begin{align}
    q\,\Delta V^{\mathrm{th}}(v_\mathrm{m})
    &=
    m e^{-\omega_\mathrm{m}}
    \left[
    1+\cosh(\omega_\mathrm{g}-\omega_\mathrm{m})
    \right]
    \nonumber\\
    &=
    2m e^{-\omega_\mathrm{m}}
    \cosh^2\left(
    \frac{\omega_\mathrm{g}-\omega_\mathrm{m}}{2}
    \right),
    \label{eq:s_thresh_rapidity_exact}
\end{align}
where the identity
$1+\cosh x=2\cosh^2(x/2)$ has been used. This is the exact
threshold dependence, corresponding to Eq.~(8) in the main text.

In the velocity-matching limit,
$v_\mathrm{m}\to v_\mathrm{g}^{-}$ and
$\omega_\mathrm{m}\to\omega_\mathrm{g}$, so that
\begin{equation}\label{eq:qDVthvg}
    q\,\Delta V^{\mathrm{th}}(v_\mathrm{g})
    =
    2m e^{-\omega_\mathrm{g}},
\end{equation}
in agreement with
Sec.~\ref{sec:s_sub_thresh_vel_match}, while in the static limit,
$v_\mathrm{m}=0$ and $\omega_\mathrm{m}=0$,
Eq.~\eqref{eq:s_thresh_rapidity_exact} reduces to
\begin{align}
    q\,\Delta V^{\mathrm{th}}(0)
    &=
    2m\cosh^2\left(\frac{\omega_\mathrm{g}}{2}\right)
    \nonumber\\
    &=
    m(\gamma_\mathrm{g}+1)
    \nonumber\\
    &=
    E_\mathrm{i}-qV_1+m,
\end{align}
recovering the conventional static threshold.

\suppsubsubsection{Near-Matching Expansion}
\label{sec:s_sub_Near_Matching}

To determine the sensitivity to a small velocity mismatch, we define
the rapidity difference
\begin{equation}
    \Delta\omega
    \equiv
    \omega_\mathrm{g}-\omega_\mathrm{m}.
\end{equation}
Dividing Eq.~\eqref{eq:s_thresh_rapidity_exact} by Eq.~\ref{eq:qDVthvg} gives
\begin{equation}
    \frac{
    \Delta V^{\mathrm{th}}(v_\mathrm{m})
    }{
    \Delta V^{\mathrm{th}}(v_\mathrm{g})
    }
    =
    e^{\Delta\omega}
    \cosh^2\left(\frac{\Delta\omega}{2}\right).
    \label{eq:s_thresh_ratio_exact}
\end{equation}
For $|\Delta\omega|\ll1$,
\begin{equation}
    e^{\Delta\omega}
    =
    1+\Delta\omega
    +\mathcal{O}(\Delta\omega^2),
\end{equation}
whereas
\begin{equation}
    \cosh^2\left(\frac{\Delta\omega}{2}\right)
    =
    1+\mathcal{O}(\Delta\omega^2).
\end{equation}
Therefore, to the first order,
\begin{equation}
    \frac{
    \Delta V^{\mathrm{th}}(v_\mathrm{m})
    -
    \Delta V^{\mathrm{th}}(v_\mathrm{g})
    }{
    \Delta V^{\mathrm{th}}(v_\mathrm{g})
    }
    \simeq
    \Delta\omega.
    \label{eq:s_thresh_fraction_rapidity}
\end{equation}

Since
\begin{equation}
    \omega=\operatorname{arctanh}(v),
\end{equation}
its derivative is
\begin{equation}
    \frac{d\omega}{dv}
    =
    \frac{1}{1-v^2}
    =
    \gamma^2.
\end{equation}
For the one-sided velocity mismatch relevant to the overtaking
geometry,
\begin{equation}
    \delta v
    \equiv
    v_\mathrm{g}-v_\mathrm{m}>0,
\end{equation}
the rapidity difference close to velocity matching is
\begin{equation}
    \Delta\omega
    =
    \omega_\mathrm{g}-\omega_\mathrm{m}
    \simeq
    \gamma_\mathrm{g}^2\delta v.
    \label{eq:s_rapidity_velocity_mismatch}
\end{equation}
Substituting Eq.~\eqref{eq:s_rapidity_velocity_mismatch} into
Eq.~\eqref{eq:s_thresh_fraction_rapidity} gives
\begin{equation}
    \frac{
    \Delta V^{\mathrm{th}}(v_\mathrm{m})
    -
    \Delta V^{\mathrm{th}}(v_\mathrm{g})
    }{
    \Delta V^{\mathrm{th}}(v_\mathrm{g})
    }
    \simeq
    \gamma_\mathrm{g}^{2}\delta v,
    \label{eq:s_thresh_fraction_velocity}
\end{equation}
which corresponds to Eq.~(9) of the main text and is valid when
\begin{equation}
    \gamma_\mathrm{g}^{2}\delta v\ll1.
\end{equation}
Consequently, restricting the fractional threshold increase to
$\epsilon$ requires approximately
\begin{equation}
    \delta v
    \lesssim
    \frac{\epsilon}{\gamma_\mathrm{g}^{2}}.
\end{equation}

\end{document}